\documentclass[12pt,prd,tightenlines,nofootinbib,showpacs,showkeys]{revtex4}
\newcommand{\be}{\begin{equation}}
\newcommand{\ee}{\end{equation}}
\newcommand{\bdis}{\begin{displaymath}}
\newcommand{\edis}{\end{displaymath}}
\newcommand{\bga}{\begin{equation}\begin{gathered}}
\newcommand{\ega}{\end{gathered}\end{equation}}
\usepackage{bm}
\usepackage{graphics}
\usepackage{rotating}
\usepackage{epsfig}
\usepackage{amssymb}
\usepackage{amsmath}
\usepackage{amsthm}
\usepackage{amsfonts}
\usepackage{amssymb}
\usepackage{braket}
\usepackage{graphics}
\usepackage{rotating}
\usepackage{epsfig}
\usepackage{indentfirst}
\usepackage{epstopdf}

\begin{document}
\title{Hyperfine structure of P states in muonic ions of lithium, beryllium and boron}
\author{\firstname{A.~E.} \surname{Dorokhov}}
\affiliation{Joint Institute of Nuclear Research, BLTP, 141980, Moscow region,
Dubna, Russia}
\author{\firstname{A.~P.} \surname{Martynenko}}
\affiliation{Samara National Research University, Moskovskoye Shosse 34, 443086, Samara, Russia}
\author{\firstname{F.~A.} \surname{Martynenko}}
\author{\firstname{O.~S.} \surname{Sukhorukova}}
\affiliation{Samara National Research University, Moskovskoye Shosse 34, 443086, Samara, Russia}

\begin{abstract}
We calculate hyperfine structure intervals $\Delta E^{hfs}(2P_{1/2})$ and $\Delta E^{hfs}(2P_{3/2})$
for P-states in muonic ions of lithium, beryllium and boron. To construct the particle interaction
operator in momentum space we use the tensor method of projection operators on states with definite 
quantum numbers of total atomic momentum $F$ and total muon momentum $j$.
We take into account vacuum polarization, relativistic,
quadruple and structure corrections of orders $\alpha^4$, $\alpha^5$ and $\alpha^6$.
The obtained numerical values of hyperfine splittings can be used for a comparison  with future
experimental data of the CREMA collaboration.
\end{abstract}



\maketitle

\section{Introduction}

Quantum electrodynamics of the bound states is one of the most successful theories in modern physics
which was checked by means of very precise measurements for many bound states.
The current experimental program of CREMA (Charge Radius Experiments with Muonic Atoms)
collaboration is directed to study of fine and hyperfine energy structure of simple muonic atoms
\cite{crema2010,crema2013,crema2016}. It is realised successfully beginning with 2010, when
two transition frequencies
$(2S^{F=1}_{1/2}-2P^{F=2}_{3/2})$ and $(2S^{F=0}_{1/2}-2P^{F=1}_{3/2})$ in muonic hydrogen were measured 
with the accuracy which allowed to obtain one order more precise value of proton charge radius.
The measurement of three transition frequencies 
$(2S^{F=3/2}_{1/2}-2P^{F=5/2}_{3/2})$,
$(2S^{F=1/2}_{1/2}-2P^{F=3/2}_{3/2})$, $(2S^{F=1/2}_{1/2}-2P^{F=1/2}_{3/2})$
in muonic deuterium by means of laser spectroscopy
methods was also performed and gave new more precise value of deuteron charge radius.
The laser spectroscopy provides unique possibility for the check and further development 
the theoretical models connected with the investigation of fundamental structure of matter.
Experimental investigations with muonic hydrogen gave unexpected results,
discovered the problem which was called the proton radius puzzle \cite{pohl2013,crema1,pohl2016}, 
raised the question of a more accurate study of the effects of the nucleus structure \cite{bacca}.
This problem remains unsolved
for a long time while there were many efforts to explain the difference of proton charge radii 
extracted from electronic and muonic atoms. Among recent papers devoted to the determination
of the proton charge radius it is necessary to mention experimental work \cite{bezginov}
where the proton charge radius is extracted from the Lamb shift measurement in ordinary hydrogen
with the value $r_p=(0.833\pm 0.010)~fm$ which is very close to the result obtained by CREMA collaboration.
Another recent work \cite{alarcon} on extraction the proton charge radius from the elastic form factor (FF) data
gave the value $r_p=0.844(7)~fm$ which is consistent with the high-precision muonic hydrogen results.
Only in the case of simple two particle bound states
the theoretical methods are well developed to calculate the energy levels including nuclear effects
from first principles. In general, it can be noted that the program for studying muon systems 
is gaining momentum: a precision study of the ground state hyperfine structure of muonic hydrogen, 
the energy interval $(1S-2S)$, and the processes of production of dimuonium are planned 
\cite{pohl2016,pp1,ma-2017,adamczak-2017}.

One of the future scientific directions of CREMA collaboration is related with
light muonic atoms of lithium, beryllium and boron \cite{pp1}. Such experiments could complete existing data
and obtain new values of charge radii of Li, Be and B. The program of experiments with light
muonic atoms was discussed many years ago in \cite{drake1,drake2}, where the estimation 
of different energy intervals in the leading order was performed. In our recent papers \cite{apm2016,apm2018}
we calculated some corrections to the Lamb shift (2P-2S) and hyperfine splitting of S-states in
muonic lithium, beryllium and boron and obtain more precise values of these energy intervals.
This work continues our investigation in \cite{apm2016,apm2018} to the case of P-wave part of the
spectrum.
The account of hyperfine structure of P-levels is also necessary because experimental transition
frequencies are measured between different components of 2P and 2S states. Despite the fact that
general theoretical methods for the calculation of hyperfine structure of P-states are well developed,
specific numerical calculations of different hyperfine structure intervals in the case of P-states
in muonic lithium, beryllium and boron were not considered in detail. So, the work \cite{drake1}
contains only general formula of hyperfine structure in leading order. Numerical calculation of HFS
in muonic Li, Be, B represents both pure theoretical interest because we have in this task
nuclei with spins different from 1/2, and experimental interest connected with experiments
of CREMA collaboration. Therefore, the aim of this work is to calculate hyperfine splitting intervals
for P-states in muonic Li, Be, B with the account of corrections to vacuum polarization and nuclear
structure.

\section{General formalism}

Among nuclei of Li, Be, B there are nuclei of different spins. For the nuclei with spin $I=s_2=1$ our
approach to the calculation of hyperfine structure was developed in \cite{apm2015} by the example
of muonic deuterium. All nuclei of lithium, beryllium and boron have isotopes with spin $s_2=3/2$
so we describe further hyperfine structure of such muonic ions which consists of six states:
$2^3P_{1/2}$, $2^5P_{1/2}$, $2^1P_{3/2}$, $2^3P_{3/2}$, $2^5P_{3/2}$, $2^7P_{3/2}$,
where the lower index corresponds to muon total momentum ${\bf j}={\bf s}_1+{\bf L}$ and
upper index is the factor $(2F+1)$ (${\bf F}={\bf j}+{\bf s}_2)$.
The contribution of the leading order $\alpha^4$ to HFS of P-states is determined by the amplitude
of one-photon interaction which is denoted $T_{1\gamma}$. 
In this work, we use both the momentum and coordinate representations to describe the interaction 
of particles. We find it useful to begin with momentum
representation of interaction amplitude in which the two-particle bound state
wave function of 2P-state can be written in the tensor form:
\begin{equation}
\label{eq1}
\psi_{2P}({\bf p})=\left(\varepsilon\cdot n_p\right)R_{21}(p),
\end{equation}
where $\varepsilon_\delta$ is the polarization vector of orbital motion, $n_p=(0,{\bf p}/p)$, $R_{21}(p)$ is
the radial wave function in momentum space. Then the contribution to the energy spectrum is determined
by the integral:
\begin{equation}
\label{eq2}
\Delta E^{hfs}=\int\left(\varepsilon^\ast\cdot n_q\right)R_{21}(q)\frac{d{\bf q}}{(2\pi)^{3/2}}
\int\left(\varepsilon\cdot n_p\right)R_{21}(p)\frac{d{\bf p}}{(2\pi)^{3/2}} \Delta V^{hfs}({\bf p},{\bf q}).
\end{equation}
The hyperfine potential $\Delta V^{hfs}$ can be constructed by means of one-photon interaction amplitude $T_{1\gamma}$
using the method of projection operators on states with definite quantum numbers 
\cite{apm2015,apm2014,koerner,berends,kuhn,grotch}. 
These projection operators
can be written in terms of particle wave functions at the rest frame in covariant form.
They allow us to avoid direct cumbersome multiplication of different factors
in the amplitudes and use the computer methods for calculating amplitudes and the energy levels
\cite{form}.
For their construction we use two different schemes
of momentum adding: 1. ${\bf J}={\bf s}_1+{\bf L}$, ${\bf F}={\bf J}+{\bf s}_2$, 
2. ${\bf S}={\bf s}_1+{\bf s}_2$, ${\bf F}={\bf S}+{\bf L}$. Taking into account that the
nuclei with spin 3/2 are described in the Rarita-Schwinger formalism by the spin-vector $v_{\alpha}(p)$
we can write the one-photon interaction amplitude in the form:
\begin{equation}
\label{eq3}
T_{1\gamma}({\bf p},{\bf q})=4\pi Z\alpha \left(\varepsilon^\ast\cdot n_q\right)\left[\bar u(q_1)
\left(\frac{(p_1+q_1)_\mu}{2m_1}+(1+a_\mu)\sigma_{\mu\epsilon}\frac{k_\epsilon}{2m_1}\right)u(p_1)\right]
\left(\varepsilon\cdot n_p\right)D_{\mu\nu}(k)\times
\end{equation}
\begin{displaymath}
\bar v_\alpha(p_2)\Bigl\{g_{\alpha\beta}\frac{(p_2+q_2)_\nu}{2m_2}F_1(k^2)+g_{\alpha\beta}\sigma_{\nu\lambda}
\frac{k^\lambda}{2m_2}F_2(k^2)+
\end{displaymath}
\begin{displaymath}
\frac{k_\alpha k_\beta}{4m_2^2}\frac{(p_2+q_2)_\nu}{2m_2}F_3(k^2)+
\frac{k_\alpha k_\beta}{4m_2^2}\sigma_{\nu\lambda}\frac{k^\lambda}{2m_2}F_4(k^2)\Bigr\}
v_\beta(q_2),
\end{displaymath}
where $p_{1,2}=\frac{m_{1,2}}{(m_1+m_2)}P\pm p$ are four-momenta of initial muon and nuclear, $q_{1,2}=
\frac{m_{1,2}}{(m_1+m_2)}Q\pm q$
are four-momenta of final muon and nuclear, $a_\mu$ is the muon anomalous magnetic moment. 
They are expressed in terms of total two-particle momenta $P,Q$
and relative momenta $p,q$.
$D_{\mu\nu}(k)$ is the photon propagator which is taken to be in the Coulomb gauge.
Four form factors which parametrise the nucleus electromagnetic current can be expressed through
multipole form factors measured in experiments:
charge $G_{E0}$, electroquadrupole $G_{E2}$, magnetic-dipole $G_{M1}$ and
magnetic-octupole  $G_{M3}$ form factors. Corresponding equations are the following \cite{nozawa,aliev,deser}:
\begin{equation}
\label{eq4}
G_{E0}(k^2)=(1+\frac{2}{3}\tau)[F_1+\tau(F_1-F_2)]-\frac{1}{3}\tau(1+\tau)[F_3+\tau(F_3-F_4)],
\end{equation}
\begin{equation}
\label{eq5}
G_{E2}(k^2)=[F_1+\tau(F_1-F_2)]-\frac{1}{2}(1+\tau)[F_3+\tau(F_3-F_4)],
\end{equation}
\begin{equation}
\label{eq6}
G_{M1}(k^2)=(1+\frac{4}{5}\tau)F_2-\frac{2}{5}\tau(1+\tau)F_4,
\end{equation}
\begin{equation}
\label{eq7}
G_{M3}(k^2)=F_2-\frac{1}{2}(1+\tau)F_4,
\end{equation}
where $\tau=-k^2/4m_2^2$. The calculation of these form factors can not be carried out with high
accuracy in nuclear models. So, we can use for them experimental data suggesting that they can be
improved if necessary.
The nucleus spin-vector wave function is described by
\begin{equation}
\label{eq8}
\psi_{\mu,\alpha}({\bf p},\sigma)=\sum_{\lambda,\omega}\Braket{\frac{1}{2}\omega;1\lambda|\frac{3}{2}\sigma}\varepsilon_\mu({\bf p},\lambda)
u_\alpha({\bf p},\omega),
\end{equation}
where $\Braket{\frac{1}{2}\omega;1\lambda|\frac{3}{2}\sigma}$ are the Clebsch-Gordan coefficients.
In order to obtain the contribution to HFS of order $\alpha^4$ including recoil effects we
should know also the transformation law of spin-vector to the rest frame.
Explicit expression of the transformation law of vector wave function $\varepsilon_{\sigma}(p)$
has the form:
\begin{equation}
\label{eq9}
\varepsilon_{\sigma}(p_2)=\varepsilon_{\sigma}(0)-\frac{p_{2,\sigma}+g_{0\sigma}m_2}{\epsilon_2(p)+m_2}\frac{(\varepsilon_{\sigma}(0)\cdot p_2)}{m_2}.
\end{equation}

\begin{table}[htbp]
\caption{The nucleus parameters of lithium, beryllium and boron.}
\label{tb1}
\bigskip
\begin{ruledtabular}
\begin{tabular}{|c|c|c|c|c|c|c|}   \hline
Nucleus & Spin &  Mass ,    & Magnetic dipole  & Charge radius, & Electroquadrupole  &  Magnetic octupole \\  
   &     & GeV     &   moment, nm   &   fm    & moment, fm$^2$   & moment, nm$\cdot$fm$^2$   \\    \hline
$^7_3Li$ & 3/2 & 6.53383 & 3.256427(2) & $2.4440\pm 0.0420 $  & -4.06(8) &  7.5 \\ \hline
$^9_4Be$ & 3/2 & 8.39479 & -1.177432(3) & $2.5190\pm 0.0120 $ & 5.29(4)  &  4.1  \\ \hline
$^{11}_5B$ & 3/2 & 10.25510 & 2.6886489(10) & $2.4060\pm 0.0294 $ &  4.07(3)& 7.8   \\ \hline
\end{tabular}
\end{ruledtabular}
\end{table}
 
To describe the hyperfine structure of state $2P_{1/2}$ we introduce in \eqref{eq3} on the first stage
of the transformation the projection operator on the muon state with j=1/2:
\begin{equation}
\label{eq10}
\hat\Pi_{j=1/2}=[u(0)\varepsilon_\omega(0)]_{j=1/2}=\frac{1}{\sqrt{3}}\gamma_5(\gamma_\omega-v_\omega)\psi(0),
\end{equation}
where $\psi(0)$ is the Dirac spinor describing the muon state with j=1/2, $v=(1,0,0,0)=P/(m_1+m_2)$ is
the auxiliary four vector.
On the second stage we
should project muon-nucleus pair on state with total momentum $F=2$ or $F=1$. In the case of state with $F=2$
the projection operator has the form
\begin{equation}
\label{eq11}
\hat\Pi_{j=1/2}(F=2)=[\psi(0)\bar v_\alpha(0)]_{F=2}=\frac{1+\hat v}{2\sqrt{2}}\gamma_\tau\varepsilon_{\alpha\tau},
\end{equation}
where the tensor $\varepsilon_{\alpha\tau}$ describes the state $F=2$. For a construction of the muon-nucleus 
interaction operator in this state
we make the summation over projections of the total momentum $F$ using the equation
\begin{equation}
\label{eq12}
\sum_{{M_F}=-2}^{2} 
\varepsilon^\ast_{\beta\lambda}\varepsilon_{\alpha\rho}=\hat\Pi_{\beta\lambda,\alpha\rho}=
[\frac{1}{2}X_{\beta\alpha}X_{\lambda\rho}+\frac{1}{2}X_{\beta\rho}X_{\lambda\alpha}-
\frac{1}{3}X_{\beta\lambda}X_{\alpha\rho}],~~~X_{\beta\alpha}=(g_{\alpha\beta}-v_\beta v_\alpha).
\end{equation}
Then the averaged over the projections $M_F$ amplitude takes the form:
\begin{equation}
\label{eq13}
\overline{T_{1\gamma}({\bf p},{\bf q})}^{j=1/2}_{F=2}=\frac{Z\alpha}{5} n_q^\delta n_p^\omega
Tr\Bigl\{\gamma_\sigma\frac{1+\hat v}{2\sqrt{2}}(\gamma_\delta-v_\delta)\gamma_5\frac{(\hat q_1+m_1)}{2m_1}
\Gamma_\mu
\frac{(\hat p_1+m_1)}{2m_1}\gamma_5(\gamma_\omega-v_\omega)\times
\end{equation}
\begin{equation*}
\frac{1+\hat v}{2\sqrt{2}}\gamma_\rho
\frac{(\hat p_2-m_2)}{2m_2}\Gamma_{\alpha\beta}^\nu\frac{(\hat q_2-m_2)}{2m_2}
\Bigr\}D_{\mu\nu}(k)\hat\Pi_{\beta_1\sigma,\alpha_1\rho}L_{\alpha\alpha_1}L_{\beta\beta_1},
\end{equation*}
where we introduce for the convenience short designations of nucleus vertex function
\begin{equation}
\label{eq14}
\Gamma_{\alpha\beta}^\nu=\Bigl[g_{\alpha\beta}\frac{(p_2+q_2)_\nu}{2m_2}F_1(k^2)+g_{\alpha\beta}\sigma_{\nu\lambda}
\frac{k^\lambda}{2m_2}F_2(k^2)+
\frac{k_\alpha k_\beta}{4m_2^2}\frac{(p_2+q_2)_\nu}{2m_2}F_3(k^2)+
\frac{k_\alpha k_\beta}{4m_2^2}\sigma_{\nu\lambda}\frac{k^\lambda}{2m_2}F_4(k^2)\Bigr],
\end{equation}
the lepton vertex function
\begin{equation}
\label{eq15}
\Gamma_\mu=\frac{p_{1,\mu}+q_{1,\mu}}{2m_1}+(1+a_\mu)\sigma_{\mu\epsilon}\frac{k_\epsilon}{2m_1}
\end{equation}
and the Lorentz factors of vector fields
\begin{equation}
\label{eq16}
L_{\alpha\alpha_1}L_{\beta\beta_1}=\left[g_{\alpha\alpha_1}-(v_\alpha-\frac{p_\alpha}{2m_2})(v_{\alpha_1}-\frac{p_{\alpha_1}}{m_2})\right]
\left[g_{\beta\beta_1}-(v_{\beta}-\frac{p_{\beta}}{2m_2})(v_{\beta_1}-\frac{p_{\beta_1}}{m_2})\right].
\end{equation}
We introduce in \eqref{eq13}
the factor $3/4\pi$ connected with the normalization condition of polarization vector in \eqref{eq1}.
The remaining cumbersome part of calculating the trace and numerous convolutions by the Lorentz indices 
can be performed by means of the package Form \cite{form}. As a result we obtain the muon-nucleus interaction
operator for the state $2^5P_{1/2}$ in the form:
\begin{equation}
\label{eq17}
V_{1\gamma}({\bf p},{\bf q})_{F=2}^{j=1/2}=\frac{2\alpha \mu_N}{27m_1m_p({\bf p}-{\bf q})^2}\Bigl\{
\frac{9}{2}pq+\frac{9m_1}{4m_2}pq-\frac{9}{4}({\bf p}{\bf q})\left(\frac{p}{q}+\frac{q}{p}\right)
+\frac{27m_1}{8m_2}({\bf p}{\bf q})\left(\frac{p}{q}+\frac{q}{p}\right)
\end{equation}
\begin{displaymath}
-\frac{9m_1}{m_2}\frac{({\bf p}{\bf q})^2}{pq}+a_\mu\Bigl[\frac{9}{4}pq-
\frac{9}{4}({\bf p}{\bf q})\left(\frac{p}{q}+\frac{q}{p}\right)
+\frac{9}{4}\frac{({\bf p}{\bf q})^2}{pq}\Bigr]+\frac{a_\mu}{F_2(0)}\Bigl[
-\frac{27}{4}pq\left(1+\frac{m_2}{m_1}\right)+
\frac{27}{4}\frac{({\bf p}{\bf q})^2}{pq}+
\end{displaymath}
\begin{displaymath}
\frac{27m_2}{8m_1}({\bf p}{\bf q})\left(\frac{p}{q}+\frac{q}{p}\right)
\Bigr]+\frac{27({\bf p}{\bf q})({\bf p}^2-{\bf q}^2)^2}{8({\bf p}-{\bf q})^2F_2(0)pq}-\frac{27}{8F_2(0)}
\Bigl[pq(2+\frac{m_1}{m_2}+\frac{m_2}{m_1})-
\frac{m_2}{m_1}({\bf p}{\bf q})\left(\frac{p}{q}+\frac{q}{p}\right)
\end{displaymath}
\begin{displaymath}
+({\bf p}{\bf q})\left(\frac{p}{q}+\frac{q}{p}\right)+4m_1m_2\frac{({\bf p}{\bf q})}{pq}-\frac{2m_1}{m_2}
\frac{({\bf p}{\bf q})^2}{pq}\Bigr]\Bigr\},
\end{displaymath}
where we set $F_1(0)=1$.
This expression clearly shows the general structure of potentials for various states, 
which we obtain at the exit from Form. Typical momentum integrals that must be calculated 
in the hyperfine structure \eqref{eq2} have the form:
\begin{equation}
\label{eq18}
J_1=\int R_{21}(q)\frac{d{\bf q}}{(2\pi)^{3/2}}
\int R_{21}(p)\frac{d{\bf p}}{(2\pi)^{3/2}} \frac{pq}{({\bf p}-{\bf q})^2}=\Braket{\frac{pq}{({\bf p}-{\bf q})^2}}
=\frac{3}{16},
\end{equation}
\begin{equation*}
J_2=\Braket{\frac{({\bf p}{\bf q})^2}{pq({\bf p}-{\bf q})^2}}=\frac{5}{48},~J_3=\Braket{\frac{({\bf p}{\bf q})(p^2+q^2)}{pq({\bf p}-{\bf q})^2}}=\frac{5}{24},~J_4=
\Braket{\frac{({\bf p}{\bf q})({\bf p}^2-{\bf q}^2)^2}{pq({\bf p}-{\bf q})^4}}=\frac{1}{6}.
\end{equation*}
It is important to note that when constructing potentials in this way, we obtain not only the hyperfine 
part of the potentials, but also the Coulomb contributions and contributions to the fine structure, 
which are further reduced when considering hyperfine splitting. Let us consider also the construction
of the potential in the case of $2^3P_{1/2}$ state. To introduce projection operators for the state 
$F = 1$, $j = 1/2$, it is necessary to add the spin of the nucleus $s_2=3/2$ and the total moment 
of the muon $j=1/2$. For this we use a basis transformation of the following form:
\begin{equation}
\label{eq19}
\Psi_{s_2=3/2,F=1,M_F}=\sqrt{\frac{2}{3}}\Psi_{\tilde S=0,F=1,M_F}+\sqrt{\frac{1}{3}}\Psi_{\tilde S=1,F=1,M_F},
\end{equation}
where the state with $s_2 = 3/2$ is represented as the sum of two moments $\tilde s_2 = 1/2$ and $l_2 = 1$,
$\tilde S=s_1+\tilde s_2$. Further, when working with $\Psi_{\tilde S=0,F=1,M_F}$ and $\Psi_{\tilde S=1,F=1,M_F}$
we introduce projection operators on these states, the form of which is well known:
\begin{equation}
\label{eq20}
\hat\Pi_\alpha(\tilde S=0,F=1)=\frac{1+\hat v}{2\sqrt{2}}\gamma_5\varepsilon_\alpha,~
\hat\Pi_\alpha(\tilde S=1,F=1)=\frac{1+\hat v}{4}\gamma_\sigma\varepsilon_{\alpha\sigma\rho\omega}v^\rho\varepsilon^\omega,
\end{equation}
where the polarization vector $\varepsilon^\omega$ in right part of \eqref{eq20} describes the state with $F=1$.
When using expansion \eqref{eq16}, several contributions to the interaction potential of particles 
in the state $F = 1$ arise, which are determined by two expressions for $\tilde S=0$ and $\tilde S=1$
with weight factors 2/3 and 1/3 respectively:
\begin{equation}
\label{eq21}
\overline{T_{1\gamma}({\bf p},{\bf q})}^{j=1/2}_{F=1}(\tilde S=0)=\frac{Z\alpha}{3} n_q^\delta n_p^\omega
Tr\Bigl\{\gamma_5\frac{1+\hat v}{2\sqrt{2}}(\gamma_\delta-v_\delta)\gamma_5\frac{(\hat q_1+m_1)}{2m_1}
\Gamma_\mu
\frac{(\hat p_1+m_1)}{2m_1}\gamma_5(\gamma_\omega-v_\omega)\times
\end{equation}
\begin{equation*}
\frac{1+\hat v}
{2\sqrt{2}}\gamma_5
\frac{(\hat p_2-m_2)}{2m_2}
\Bigl[g_{\alpha\beta}\frac{(p_2+q_2)_\nu}{2m_2}F_1(k^2)+g_{\alpha\beta}\sigma_{\nu\lambda}
\frac{k^\lambda}{2m_2}F_2(k^2)+
\frac{k_\alpha k_\beta}{4m_2^2}\frac{(p_2+q_2)_\nu}{2m_2}F_3(k^2)+
\end{equation*}
\begin{equation*}
\frac{k_\alpha k_\beta}{4m_2^2}\sigma_{\nu\lambda}\frac{k^\lambda}{2m_2}F_4(k^2)\Bigr]\frac{(\hat q_2-m_2)}{2m_2}
\Bigr\}D_{\mu\nu}(k)\hat\Pi_{\beta_1\sigma\alpha_1\rho}L_{\alpha\alpha_1}L_{\beta\beta_1}
(g_{\alpha_1\beta_1}-v_{\alpha_1}v_{\beta_1}),
\end{equation*}
\begin{equation}
\label{eq22}
\overline{T_{1\gamma}({\bf p},{\bf q})}^{j=1/2}_{F=1}(\tilde S=1)=\frac{Z\alpha}{3} n_q^\delta n_p^\omega
Tr\Bigl\{\gamma_\rho\frac{1+\hat v}{4}(\gamma_\delta-v_\delta)\gamma_5\frac{(\hat q_1+m_1)}{2m_1}
\Gamma_\mu
\frac{(\hat p_1+m_1)}{2m_1}\gamma_5(\gamma_\omega-v_\omega)\times
\end{equation}
\begin{equation*}
\frac{1+\hat v}{4}\gamma_\tau\frac{(\hat p_2-m_2)}{2m_2}
\Bigl[g_{\alpha\beta}\frac{(p_2+q_2)_\nu}{2m_2}F_1(k^2)+g_{\alpha\beta}\sigma_{\nu\lambda}
\frac{k^\lambda}{2m_2}F_2(k^2)+
\frac{k_\alpha k_\beta}{4m_2^2}\frac{(p_2+q_2)_\nu}{2m_2}F_3(k^2)+
\end{equation*}
\begin{equation*}
\frac{k_\alpha k_\beta}{4m_2^2}\sigma_{\nu\lambda}\frac{k^\lambda}{2m_2}F_4(k^2)\Bigr]\frac{(\hat q_2-m_2)}{2m_2}
\Bigr\}D_{\mu\nu}(k)L_{\alpha\alpha_3}L_{\beta\beta_3}
\epsilon_{\rho\beta_3\alpha_1\beta_1}\epsilon_{\tau\alpha_3\rho_1\omega_1}
(g_{\omega_1\beta_1}-v_{\omega_1}v_{\beta_1}).
\end{equation*}
There is also off-diagonal element of the form 
$<\Psi_{\tilde S=0,F=1,M_F}|V^{j=1/2}_{1\gamma,F=1}|\Psi_{\tilde S=1,F=1,M_F}>$. Omitting other details of
the calculation and using \eqref{eq17}, \eqref{eq21},
\eqref{eq22} we obtain the hyperfine splitting of $2P_{1/2}$ state as follows:
\begin{equation}
\label{eq23}
\Delta E^{hfs}(2^5P_{1/2}-2^3P_{1/2})=
\frac{2\alpha(Z\alpha)^3\mu^3\mu_N}{27m_1m_p}\bigl[1+\frac{1}{2}a_\mu+\frac{m_1}{2m_2}-
\frac{3m_1}{4m_2F_2(0)}\bigr]=
\begin{cases}
_3^7Li:~210.8960~meV,\\
_4^9Be:-183.2929~meV,\\
_5^{11}B:~818.1086~meV,
\end{cases}
\end{equation}
where we take into account only leading order terms including recoil correction related with
nucleus magnetic form factor. At $k^2=0$ we have $F_2(0)=m_2\mu_N/Zm_p$.

The calculation of hyperfine splitting for the $2^{(2F+1)}P_{3/2}$ state is a more complicated problem, 
since it is more complicated to construct projection operators for these states. 
The most simple form is the projection operator on the state with $F = 3$. In this case, 
when the two moments 3/2 are added, we get the state with the maximum total momentum, 
which is described by the tensor $\varepsilon_{\alpha\beta\gamma}$. 
The projection operator on this state is equal
\begin{equation}
\label{eq24}
\hat\Pi_{\omega\alpha_1}=[u_\omega(0)\bar v_{\alpha_1}(0)]_{F=3}^{j=3/2}=
\frac{1+\hat v}{2\sqrt{2}}\gamma_{\omega_1}\varepsilon_{\omega\alpha_1\omega_1},
\end{equation}
and a summation over projections has the form \cite{grotch}:
\begin{equation}
\label{eq25}
\sum_{{M_F}=-3}^{3} 
\varepsilon^\ast_{\alpha\omega\omega_1}\varepsilon_{\beta\delta\delta_1}=
\hat\Pi_{\alpha\omega\omega_1,\beta\delta\delta_1}=
\left[\frac{1}{6}\Omega^{(1)}_{\alpha\omega\omega_1,\beta\delta\delta_1}-
\frac{1}{15}\Omega^{(2)}_{\alpha\omega\omega_1,\beta\delta\delta_1}\right],
\end{equation}
\begin{equation*}
\Omega^{(1)}_{\alpha\omega\omega_1,\beta\delta\delta_1}=X_{\alpha\beta}X_{\omega\delta}X_{\omega_1\delta_1}+
X_{\alpha\beta}X_{\omega\delta_1}X_{\omega_1\delta}+
X_{\alpha\delta}X_{\omega\beta}X_{\omega_1\delta_1}+
\end{equation*}
\begin{equation*}
X_{\alpha\delta}X_{\omega\delta_1}X_{\omega_1\beta}+
X_{\alpha\delta_1}X_{\omega\delta}X_{\omega_1\beta}+
X_{\alpha\delta_1}X_{\omega\beta}X_{\omega_1\delta},
\end{equation*}
\begin{equation*}
\Omega^{(2)}_{\alpha\omega\omega_1,\beta\delta\delta_1}=
X_{\alpha\omega}X_{\omega_1\delta_1}X_{\beta\delta}+
X_{\alpha\omega}X_{\omega_1\delta}X_{\beta_1\delta_1}+
X_{\alpha\omega}X_{\omega_1\beta}X_{\delta\delta_1}+
X_{\alpha\omega_1}X_{\omega\delta_1}X_{\beta\delta}+
\end{equation*}
\begin{equation*}
X_{\alpha\omega_1}X_{\omega\delta}X_{\beta\delta_1}+
X_{\alpha\omega_1}X_{\omega\beta}X_{\delta\delta_1}+
X_{\omega\omega_1}X_{\alpha\delta_1}X_{\beta\delta}+
X_{\omega\omega_1}X_{\alpha\delta}X_{\beta\delta_1}+
X_{\omega\omega_1}X_{\alpha\beta}X_{\delta\delta_1}.
\end{equation*}
Then the interaction amplitude averaged over the projections $M_F$ can be represented as follows:
\begin{equation}
\label{eq26}
\overline{T_{1\gamma}({\bf p},{\bf q})}^{j=3/2}_{F=3}=\frac{3Z\alpha}{7} n_q^\delta n_p^\omega
Tr\Bigl\{\gamma_{\delta_1}\frac{1+\hat v}{2\sqrt{2}}\frac{(\hat q_1+m_1)}{2m_1}
\Gamma_\mu
\frac{(\hat p_1+m_1)}{2m_1}\frac{1+\hat v}{2\sqrt{2}}\gamma_{\omega_1}\times
\end{equation}
\begin{equation*}
\frac{(\hat p_2-m_2)}{2m_2}\Gamma_{\alpha\beta}^\nu\frac{(\hat q_2-m_2)}{2m_2}
\Bigr\}D_{\mu\nu}(k)L_{\alpha\alpha_1}L_{\beta\beta_1}\hat\Pi_{\beta_1\delta\delta_1,\alpha_1\omega\omega_1}.
\end{equation*}

To calculate the interval of the hyperfine structure $\Delta E^{hfs}(2^7P_{3/2}-2^5P_{3/2})$, 
it is also necessary to build the potential 
for the state with $F = 2$, which is obtained by adding two 3/2 moments. Acting as in \eqref{eq19}, we first 
represent the state of the nucleus as the result of adding the two moments $\tilde s_2=1/2$ and $l_2=1$
and introduce the momentum ${\bf j}_1 = {\bf j} + \tilde {\bf s}_2$, which takes values 2 and 1:
\begin{equation}
\label{eq27}
\Psi_{s_2=3/2,j=3/2,F=2}=\frac{1}{\sqrt{2}}\Psi_{j_1=2,l_2=1,F=2}+\frac{1}{\sqrt{2}}\Psi_{j_1=1,l_2=1,F=2}.
\end{equation}
The projection operator on the state $j_1 = 2$ has the form:
\begin{equation}
\label{eq28}
\hat\Pi_{\alpha,j_1=2}^{j=3/2}=[u_\alpha(0)\bar v(0)]^{j=3/2}_{j_1=2}=\frac{1+\hat v}{2\sqrt{2}}\gamma^{\beta_1}
\varepsilon_{\alpha\beta_1}.
\end{equation}
In order to write a projection operator on the state $j_1 = 1$, we will already represent 
the moment of the muon $j=3/2$ as the result of adding two moments: ${\bf j}={\bf s}_1+{\bf l}_1$. 
Given the coefficients  of vector addition of moments, we obtain the following expansion:
\begin{equation}
\label{eq29}
\Psi_{s_2=3/2,j=3/2,F=2}=\frac{1}{\sqrt{2}}\Psi_{j_1=2,l_2=1,F=2}+
\frac{1}{\sqrt{3}}\Psi_{(S=0,l_1=1,j_1=1),l_2=1,F=2}+
\frac{1}{\sqrt{6}}\Psi_{(S=1,l_1=1,j_1=1),l_2=1,F=2}.
\end{equation}
Performing the addition of individual moments in $F = 2$, we obtain the following result 
for the projection operator on a state with $F = 2$:
\begin{equation}
\label{eq30}
\hat\Pi^{j=3/2}_{\alpha\beta,F=2}=\frac{(1+\hat v)}{2\sqrt{6}}\gamma_5\varepsilon_{\alpha\beta}+
\frac{(1+\hat v)}{4\sqrt{6}}[-g_{\gamma\alpha}\epsilon_{\beta\alpha_1\alpha_2\alpha_3}+
g_{\gamma\beta}\epsilon_{\alpha\alpha_1\alpha_2\alpha_3}+g_{\gamma\alpha_1}\epsilon_{\alpha\beta\alpha_2\alpha_3}]
v^{\alpha_2}\gamma^{\alpha_1}\varepsilon^{\gamma\alpha_3}.
\end{equation}
As we see, the tensor in the right-hand side of \eqref{eq29} contains both the symmetric 
and antisymmetric parts in the indices $\alpha$, $\beta$.
The same decomposition and addition of individual moments is also used to obtain projection 
operators on other states of the hyperfine structure with $F = 0$ and $F = 1$. They have the form:
\begin{equation}
\label{eq31}
\hat\Pi^{j=3/2}_{F=0}=[u_\omega\bar v_\alpha(0)]^{j=3/2}_{F=0}=
\frac{1+\hat v}{6}\gamma_5(g_{\omega\alpha}-v_\omega v_\alpha)-
\frac{(1+\hat v)}{12}\gamma^\lambda\epsilon_{\lambda\omega\sigma\alpha}v^\sigma,
\end{equation}
\begin{equation}
\label{eq32}
\hat\Pi^{j=3/2}_{F=1}=[u_\omega\bar v_\alpha(0)]^{j=3/2}_{F=1}=
\frac{1+\hat v}{24\sqrt{5}}\Bigl[
-8(g_{\omega\alpha}-v_\omega v_{\alpha})\gamma_{\alpha_2}+2(g_{\omega\alpha_2}\gamma_\alpha+
g_{\alpha\alpha_2}\gamma_\omega)-
\end{equation}
\begin{equation*}
2(g_{\alpha_2\omega}v_\alpha+g_{\alpha\alpha_2}v_\omega)
-10\epsilon_{\omega\alpha\alpha_3\alpha_2}v^{\alpha_3}\gamma_5\Bigr]\varepsilon^{\alpha_2}.
\end{equation*}

In the practical use of \eqref{eq31} - \eqref{eq32} , it is convenient to single out the contributions 
of the symmetric and antisymmetric parts of the projection operators. The general structure 
of the amplitudes and interaction potentials of particles in these states has the same form 
as \eqref{eq13}, \eqref{eq17}, \eqref{eq21}, \eqref{eq22}, and the intervals of the hyperfine 
structure themselves are determined by formulas similar to \eqref{eq23}:
\begin{equation}
\label{eq33}
\Delta E^{hfs}(2^7P_{3/2}-2^5P_{3/2})=
\frac{\alpha(Z\alpha)^3\mu^3\mu_N}{45m_1m_p}\bigl[1-\frac{1}{4}a_\mu+\frac{5m_1}{4m_2}
-\frac{15m_1}{8m_2F_2(0)}\bigr]=
\begin{cases}
_3^7Li:~63.8246~meV,\\
_4^9Be:-55.7466~meV,\\
_5^{11}B:~246.6252~meV,
\end{cases}
\end{equation}
\begin{equation}
\label{eq34}
\Delta E^{hfs}(2^5P_{3/2}-2^3P_{3/2})=
\frac{2\alpha(Z\alpha)^3\mu^3\mu_N}{135m_1m_p}\bigl[1-\frac{1}{4}a_\mu+\frac{5m_1}{4m_2}-
\frac{15m_1}{8m_2F_2(0)}\bigr]=
\begin{cases}
_3^7Li:~42.5497~meV,\\
_4^9Be:-37.1644~meV,\\
_5^{11}B:~164.4168~meV,
\end{cases}
\end{equation}
\begin{equation}
\label{eq35}
\Delta E^{hfs}(2^3P_{3/2}-2^1P_{3/2})=
\frac{\alpha(Z\alpha)^3\mu^3\mu_N}{135m_1m_p}\bigl[1-\frac{1}{4}a_\mu+\frac{5m_1}{4m_2}-
\frac{15m_1}{8m_2F_2(0)}\bigr]=
\begin{cases}
_3^7Li:~21.2932~meV,\\
_4^9Be:-18.5822~meV,\\
_5^{11}B:~82.2084~meV,
\end{cases}
\end{equation}
The numerical values of the contributions \eqref{eq23}, \eqref{eq33}-\eqref{eq35} are large.
Therefore, to increase 
the accuracy of calculations, it makes sense to consider a number of corrections to these formulas
what we do below in other sections.

However, it is useful to consider another approach to solving this problem in the coordinate 
representation, which is widespread \cite{egs,sobel,borie}. Since in muon ions we encounter 
nuclei of different spins, 
it is necessary to have a two-particle Hamiltonian for electromagnetically interacting particles 
of arbitrary spin. Some time ago, the task of constructing such an effective Hamiltonian 
was solved in \cite{ibk,kp2004,kp2008,kp2010,eides2010} in connection with the calculation 
of gyromagnetic factors of bound particles. 
To calculate the hyperfine structure of the spectrum of P-levels, it is necessary to use the following 
term from this Hamiltonian:
\begin{equation}
\label{eq36}
\Delta H^{hfs}=\frac{Z\alpha g_N}{2m_1m_2r^3}\left[1+\frac{m_1}{m_2}-\frac{m_1}{m_2g_N}\right]({\bf L}{\bf s}_2)-
\frac{Z\alpha (1+a_\mu)g_N}{2m_1m_2r^3}[{\bf s}_1{\bf s}_2-3({\bf s}_1{\bf r})({\bf s}_2{\bf r})],
\end{equation}
where the gyromagnetic factor of the nucleus $g_N=F_1(0)/s_2$=$2m_2\mu_N/3Zm_p$ and nucleus magnetic
moment is taken in nuclear magnetons. To calculate the relative 
level arrangement, a fine part of the Hamiltonian is also necessary:
\begin{equation}
\label{eq37}
\Delta H^{fs}=\frac{Z\alpha}{m_1m_2r^3}\left[1+\frac{m_2}{2m_1}+a_\mu\left(1+\frac{m_2}{m_1}\right)\right]
({\bf L}{\bf s}_1).
\end{equation}
Averaging \eqref{eq37} over the wave functions of the 2P state, we obtain the main contribution to the fine splitting:
\begin{equation}
\label{eq38}
E^{fs}=\frac{(Z\alpha)^4\mu^3}{16m_1m_2}\left[1+\frac{m_2}{2m_1}+a_\mu\left(1+\frac{m_2}{m_1}\right)\right]=
\begin{cases}
_3^7Li:~747.8581~meV,\\
_4^9Be:~2372.2215~meV,\\
_5^{11}B:~5804.9674~meV,
\end{cases}
\end{equation}

While our main goal is the calculation of P-wave hyperfine splittings
we estimate here also vacuum polarization correction of order $\alpha^5$ (leading order correction)
to fine splitting. Using in this case basic relations from \cite{apm2010} for the corrections
in first order and second order perturbation theory we obtain total vacuum polarization contribution as
follows:
\begin{equation}
\label{eq38a}
\Delta E^{fs}_{vp}=
\begin{cases}
_3^7Li:~2.3483~meV,\\
_4^9Be:~10.1158~meV,\\
_5^{11}B:~30.5417~meV,
\end{cases}
\end{equation}

The hyperfine part of the Hamiltonian includes two operators
\begin{equation}
\label{eq39}
T_1={\bf L}{\bf s}_2,~~~T_2={\bf s}_1{\bf s}_2-3({\bf s}_1{\bf n})({\bf s}_2{\bf n}).
\end{equation}
Diagonal in j matrix elements contribute to the hyperfine structure in the form:
\begin{equation}
\label{eq40}
E(2^{2F+1}P_j)=\frac{\alpha(Z\alpha)^3\mu^3\mu_N}{72m_1m_p}\left[\bar{T_1}+\frac{m_1}{m_2}\bar{T_1}-
\frac{3m_1}{2m_2F_2(0)}\bar{T_1}-(1+a_\mu)\bar{T_2}\right].
\end{equation}
The calculation of matrix elements $\bar{T_1}$ and $\bar{T_2}$ is carried out using the basic formulas 
from \cite{apm2015}
(see Appendix A). As a result, the position of energy levels $2^{2F+1}P_j$
is determined by the following expressions:
\begin{equation}
\label{eq41}
E(2^7P_{3/2})=\tilde E^{fs}+\frac{\alpha(Z\alpha)^3\mu^3\mu_N}{60m_1m_p}\left[1-\frac{a_\mu}{4}+\frac{5m_1}{4m_2}-
\frac{15m_1}{8m_2F_2(0)}\right]=
\begin{cases}
_3^7Li:798.0748~meV,\\
_4^9Be:2340.5274~meV,\\
_5^{11}B:6020.4780~meV,
\end{cases}
\end{equation}
\begin{equation}
\label{eq42}
E(2^5P_{3/2})=\tilde E^{fs}-\frac{\alpha(Z\alpha)^3\mu^3\mu_N}{180m_1m_p}\left[1-\frac{a_\mu}{4}+\frac{5m_1}{4m_2}-
\frac{15m_1}{8m_2F_2(0)}\right]=
\begin{cases}
_3^7Li:734.2502~meV,\\
_4^9Be:2396.2740~meV,\\
_5^{11}B:5773.8528~meV,
\end{cases}
\end{equation}
\begin{equation}
\label{eq43}
E(2^3P_{3/2})=\tilde E^{fs}-\frac{11\alpha(Z\alpha)^3\mu^3\mu_N}{540m_1m_p}\left[1-\frac{a_\mu}{4}+\frac{5m_1}{4m_2}-
\frac{15m_1}{8m_2F_2(0)}\right]=
\begin{cases}
_3^7Li:691.7005~meV,\\
_4^9Be:2433.4383~meV,\\
_5^{11}B:5609.4360~meV,
\end{cases}
\end{equation}
\begin{equation}
\label{eq44}
E(2^1P_{3/2})=\tilde E^{fs}-\frac{\alpha(Z\alpha)^3\mu^3\mu_N}{36m_1m_p}\left[1-\frac{a_\mu}{4}+\frac{5m_1}{4m_2}-
\frac{15m_1}{8m_2F_2(0)}\right]=
\begin{cases}
_3^7Li:670.4256~meV,\\
_4^9Be:2452.0205~meV,\\
_5^{11}B:5527.2276~meV,
\end{cases}
\end{equation}
\begin{equation}
\label{eq45}
E(2^5P_{1/2})=\frac{\alpha(Z\alpha)^3\mu^3\mu_N}{36m_1m_p}\left[1+\frac{a_\mu}{2}+\frac{m_1}{2m_2}-
\frac{3m_1}{4m_2F_2(0)}\right]=
\begin{cases}
_3^7Li:79.0860~meV,\\
_4^9Be:-68.7348~meV,\\
_5^{11}B:306.7907~meV,
\end{cases}
\end{equation}
\begin{equation}
\label{eq46}
E(2^3P_{1/2})=-\frac{5\alpha(Z\alpha)^3\mu^3\mu_N}{108m_1m_p}\left[1+\frac{a_\mu}{2}+\frac{m_1}{2m_2}-
\frac{3m_1}{4m_2F_2(0)}\right]=
\begin{cases}
_3^7Li:-131.8100~meV,\\
_4^9Be:114.5581~meV,\\
_5^{11}B:-511.3179~meV,
\end{cases}
\end{equation}
where we add a sum \eqref{eq38}, \eqref{eq38a} $\tilde E^{fs} =E^{fs}+\Delta E^{fs}_{vp}$
to fix the relative position of sublevels.
The obtained expressions \eqref{eq43}-\eqref{eq46} which contain the factor $1/g_N$ 
coming from the hyperfine interaction Hamiltonian give the hyperfine splitting coinciding with
\eqref{eq23}, \eqref{eq33}-\eqref{eq35}.

\begin{table}[htbp]
\caption{Diagonal matrix elements of hyperfine structure of $2P$-states in muonic ions Li, Be, B
in first order perturbation theory.
The first, second and third lines correspond to lithium, beryllium and boron.}
\label{tb2}
\bigskip
\begin{tabular}{|c|c|c|c|c|c|c|}   \hline
The contribution & $2^3P_{1/2}$, & $2^5P_{1/2}$,& $2^1P_{3/2}$,& $2^3P_{3/2}$, & $2^5P_{3/2}$, & $2^7P_{3/2}$, \\   
   &   meV  & meV   &  meV   &  meV  & meV   & meV  \\      \hline
Leading order      &  -131.8100  &   79.0860  & 670.4256 &  691.7005 & 734.2502 &  798.0748 \\
$\alpha^4$ correction   &   114.5581  &-68.7348  &  2452.0205&    2433.4383&   2396.27400& 2340.5274\\  
   &   -511.3179  & 306.7907  &   5527.2276&   5609.4360& 5773.8528&       6020.4780\\     \hline
Quadrupole    & 0    & 0  &          -186.9598   &   -37.3920  & 112.1759 & -37.3920  \\
correction   & 0   &  0   &  583.5774   &  116.7155  &   -350.1465  & 116.7155 \\   
 of order $\alpha^4$  &   0   &   0    &   882.8935   &    176.5787   &  -529.7361  &   176.5787   \\        \hline
VP correction    & -0.5701  &  0.3420   &     -0.2784      & -0.2042  &   -0.0557   & 0.1671   \\
of order $\alpha^5$  &  0.6710   &  -0.4026   &   0.3441   &  0.2523    &  0.0689  & -0.2065 \\   
  &  -3.6909   & 2.2146   & -1.9209   & -1.4087   &  -0.3842   &  1.1526  \\      \hline
Quadrupole and  & 0   & 0   &  -0.6573         & -0.1315  & 0.3944 & -0.1315   \\
VP correction   &  0   &  0   &  2.7276  &    0.5455   & -1.6365    &  0.5455   \\
of order $\alpha^5$  &   0  &   0  &  5.0232   &   1.0046    &  -3.0140  &  1.0046   \\    \hline
Relativistic   & -0.1289   &  0.0773 &  -0.0115 &   -0.0084  & -0.0023 &  0.0069  \\
correction  & 0.1964   & -0.1178    &   0.0176  &  0.0129   & 0.0035   &  -0.0105     \\   
of order $\alpha^6$  &  -1.3686  & 0.8212   &  -0.1223   &  -0.0897   &  -0.0245  &   0.0734  \\    \hline
VP correction     &   -0.0011   & -0.0007 &   -0.0004   &  -0.0003  & -0.0001 &  0.0002     \\
of order $\alpha^6$& 0.0011  &  -0.0007  &  0.0005   & 0.0003   & 0.0001    &   -0.0003   \\   
      &     -0.0054  &    0.0032   &   -0.0023    &    -0.0017   &   -0.0005   &  0.0014  \\     \hline
Structure   &  -0.0784   & 0.0471 &  -0.0008   &    -0.0007     &  -0.0004   &  -0.0001  \\
correction   & 0.1295  &    -0.0777  &    0.0018   &     0.0015          &   0.0008     &  -0.0001  \\ 
 of order $\alpha^6$ &   -0.8292  &   0.4975  &  -0.0050   &  -0.0043  &  -0.0028  &   -0.0007  \\     \hline
Summary  &  -132.5885  & 79.5517      & 482.5174 &653.9634      & 845.9732    & 760.7254\\ 
contribution & 114.2141   & -69.3336   & 3038.6896  & 2550.9663 & 2044.5643 & 2457.5710\\ 
   &   -517.2120   &  310.3272     b & 6413.0938&  5785.8732  &  5240.6907  &  6199.2880   \\   \hline
\end{tabular}
\end{table}

\section{The contribution of quadrupole interaction}

In the leading order $\alpha^4$ in the energy spectrum of muonic ions Li, Be, B, there is another important 
contribution  of the quadrupole interaction, which must be taken into account. It arises 
for muon states with $j=3/2$ due to the fact that the nuclei have a non-spherical shape.
The calculation of this contribution to hyperfine structure in muonic
ions in coordinate space is based on the representation of quadrupole interaction as a scalar product 
of two irreducible
tensor operators of rank 2. Then the matrix elements of tensor operators are expressed in terms 
of reduced matrix elements using the Wigner-Eckart theorem \cite{sobel,apmfian}.

Using the method of projection operators formulated above, we can distinguish in the amplitude 
of the one-photon interaction a part proportional to the quadrupole form factor $G_{E2}(k^2)$. 
Its value at zero $G_{E2}(0)=m_2^2 Q/Z$, and the magnitude of the quadrupole moment of the nucleus $Q$ sets 
the numerical value of this correction.
The averaged amplitudes of quadrupole interaction for different states have the form:
\begin{equation}
\label{eq47}
\overline{T_{1\gamma,Q}({\bf p},{\bf q})}_{F=3}^{j=3/2}=\frac{\alpha Q}{20({\bf p}-{\bf q})^2}
\left[pq-4({\bf p}{\bf q})
\left(\frac{p}{q}+\frac{q}{p}\right)+7\frac{({\bf p}{\bf q})^2}{pq}\right],
\end{equation}
\begin{equation}
\label{eq48}
\overline{T_{1\gamma,Q}({\bf p},{\bf q})}_{F=2}^{j=3/2}=\frac{\alpha Q}{60({\bf p}-{\bf q})^2}
\left[9pq+4({\bf p}{\bf q})
\left(\frac{p}{q}+\frac{q}{p}\right)-17\frac{({\bf p}{\bf q})^2}{pq}\right],
\end{equation}
\begin{equation}
\label{eq49}
\overline{T_{1\gamma,Q}({\bf p},{\bf q})}_{F=1}^{j=3/2}=\frac{\alpha Q}{20({\bf p}-{\bf q})^2}
\left[pq-4({\bf p}{\bf q})
\left(\frac{p}{q}+\frac{q}{p}\right)+7\frac{({\bf p}{\bf q})^2}{pq}\right],
\end{equation}
\begin{equation}
\label{eq50}
\overline{T_{1\gamma,Q}({\bf p},{\bf q})}_{F=0}^{j=3/2}=\frac{\alpha Q}{12({\bf p}-{\bf q})^2}
\left[-3pq+4({\bf p}{\bf q})
\left(\frac{p}{q}+\frac{q}{p}\right)-5\frac{({\bf p}{\bf q})^2}{pq}\right].
\end{equation}
You may notice that the amplitudes \eqref{eq47} and \eqref{eq49} coincide. Making momentum integration
by means of \eqref{eq18} we obtain contributions to the energy levels $2^{2F+1}P_{3/2}$:
\begin{equation}
\label{eq51}
\Delta E_Q^{hfs}=\frac{\alpha Q(\mu Z\alpha)^3}{240}\left[\delta_{F3}-3\delta_{F2}+\delta_{F1}+5\delta_{F0}\right].
\end{equation}
The quadruple moments of nuclei are written in Table~\ref{tb1}.
The result \eqref{eq51} coincides exactly with previous calculations made by different method \cite{apmfian}.
Their numerical values are presented in Table~\ref{tb2}.

The magnitude of the quadrupole contribution is significant, therefore, in the case 
of quadrupole interaction, it makes sense to consider also different corrections to it. 
One of the most important effects leading to the correction of the obtained results 
in order $\alpha^4$ is the effect of vacuum polarization (VP). We begin its discussion 
precisely with the quadrupole 
interaction, since it can be simply calculated within the formulated method 
by a small modification of relations \eqref{eq47}-\eqref{eq50}.
For its calculation in momentum representation In the first order perturbation theory
we should use the following replacement in the photon propagator of \eqref{eq47}-\eqref{eq50}:
\begin{equation}
\label{eq52}
\frac{1}{({\bf p}-{\bf q})^2}\to \frac{\alpha}{3\pi}\int_1^\infty
\frac{\rho(\xi)d\xi}{({\bf p}-{\bf q})^2+4m_e^2\xi^2},~~~\rho(\xi)=\sqrt{\xi^2-1}(2\xi^2+1)/\xi^4.
\end{equation}
Then the correction to vacuum polarization in the quadrupole interaction can be expressed in terms of three
momentum integrals which are calculated analytically:
\begin{equation}
\label{eq53}
I_1=\int R_{21}(q)\frac{d{\bf q}}{(2\pi)^{3/2}}
\int R_{21}(p)\frac{d{\bf p}}{(2\pi)^{3/2}} \frac{pq}{[({\bf p}-{\bf q})^2+4m_e^2\xi^2]}=
\end{equation}
\begin{equation*}
=\Braket{\frac{pq}{[({\bf p}-{\bf q})^2+4m_e^2\xi^2]}}=\frac{a(3a+8)+6}{2(a+2)^4},~~~a=\frac{4m_e\xi}{\mu Z\alpha}.
\end{equation*}
\begin{equation*}
I_2=\Braket{\frac{({\bf p}{\bf q})^2}{pq[({\bf p}-{\bf q})^2+4m_e^2\xi^2]}}=\frac{a(3a+8)+10}{6(a+2)^4},~~~
I_3=\Braket{\frac{({\bf p}{\bf q})(p^2+q^2)}{pq[({\bf p}-{\bf q})^2+4m_e^2\xi^2]}}=\frac{2(4a+5)}{3(a+2)^4}.
\end{equation*}
After using \eqref{eq53} the energy correction becomes a function of dimensionless parameter
$a_1=\frac{4m_e}{\mu Z\alpha}$
\begin{equation}
\label{eq54}
\Delta E^Q_{vp}(2^{(2F+1)}P_{3/2})=\frac{\alpha^2(Z\alpha)^3Q}{90\pi\left(4-{a_1}^2\right)^{5/2}}
\Biggl\{2 \sqrt{4-a_1^2} \left({a_1}^2-1\right)+
\end{equation}
\begin{equation*}
\left(5{a_1}^2-8\right) \ln\Bigl[\frac{
\left(2-\sqrt{4-{a_1}^2}\right)}{{a_1} }\Bigr]\Biggr\}
\left[5\delta_{F0}+\delta_{F1}-3\delta_{F2}+\delta_{F3}\right].
\end{equation*}
This dimensionless parameter is not suitable for expansions, since $a_1(Li)=0.89795$,
$a_1(Be)=0.67109$, $a_1(B)=0.53566$.
Another contribution of VP plus quadrupole interaction of the same order $\alpha^5$ comes from
second order perturbation theory. Taking one perturbation as in \eqref{eq52} but in
coordinate representation and other perturbation as a quadrupole interaction
\begin{equation}
\label{eq55}
\Delta V_Q(r)=\frac{Z\alpha Q}{6r^3}[{\bf s}_2{\bf s}_2-3({\bf s}_2{\bf n})({\bf s}_2{\bf n})],
\end{equation}
we present necessary contribution in integral form ($b_1=2m_e/W$):
\begin{equation}
\label{eq56}
\Delta E^{hfs}_{Q,vp,sopt}=\frac{\alpha^5Z^3\mu^3Q}{144\pi}\int\limits_1^\infty\frac{\rho(\xi)d\xi}{(1+b_1)^5}
[3+11b_1+4b_1^2+4(1+b_1)\ln(1+b_1)][\frac{1}{5}\delta_{F3}-\frac{3}{5}\delta_{F2}+
\frac{1}{5}\delta_{F1}+\delta_{F0}].
\end{equation}
Corresponding numerical results of the sum of corrections \eqref{eq54} and \eqref{eq56}
for the states $2^{(2F+1)}P_{3/2}$ are included in Table~\ref{tb2}.

\section{Corrections to the vacuum polarization and nucleus structure}

The main contribution of the effects of vacuum polarization in the hyperfine structure 
of the energy spectrum of the P-states is related with a modification of the particle 
interaction potential \eqref{eq36}, which in turn is determined by the replacement \eqref{eq52}. 
Using the results of the previous section, in which the spin-orbit and spin-spin interaction 
operator is constructed in the momentum representation, we can present the corrections 
to vacuum polarization for hyperfine splitting in the form:
\begin{equation}
\label{eq57}
\Delta V^{hfs}_{vp}(2^7P_{3/2}-2^5P_{3/2})=\frac{\alpha}{135\pi}\int_1^\infty\frac{\rho(\xi)d\xi}{({\bf p}-{\bf q})^2+4m_e^2\xi^2}\Bigl\{
12pq+15\frac{m_1}{m_2}pq+12({\bf p}{\bf q})\left(\frac{p}{q}+\frac{q}{p}\right)-
\end{equation}
\begin{equation*}
36\frac{({\bf p}{\bf q})^2}{pq}-15\frac{m_1}{m_2}\frac{({\bf p}{\bf q})^2}{pq}+
a_\mu\Bigl[-3pq+12({\bf p}{\bf q})\left(\frac{p}{q}+\frac{q}{p}\right)-21\frac{({\bf p}{\bf q})^2}{pq}\Bigr]-
\frac{45m_1}{2m_2F_2(0)}\bigl[pq-\frac{{\bf p}{\bf q})^2}{pq}\bigr],
\end{equation*}
\begin{equation}
\label{eq58}
\Delta V^{hfs}_{vp}(2^5P_{3/2}-2^3P_{3/2})=\frac{2}{3}\Delta V^{hfs}_{vp}(2^7P_{3/2}-2^5P_{3/2})=
2\Delta V^{hfs}_{vp}(2^3P_{3/2}-2^1P_{3/2}).
\end{equation}
\begin{equation}
\label{eq59}
\Delta V^{hfs}_{vp}(2^5P_{1/2}-2^3P_{1/2})=\frac{2\alpha^2}{81\pi}\int_1^\infty\frac{\rho(\xi)d\xi}{({\bf p}-{\bf q})^2+4m_e^2\xi^2}\Bigl\{
12pq+6\frac{m_1}{m_2}pq-6({\bf p}{\bf q})\left(\frac{p}{q}+\frac{q}{p}\right)-
\end{equation}
\begin{equation*}
6\frac{m_1}{m_2}\frac{({\bf p}{\bf q})^2}{pq}+
a_\mu\Bigl[6pq-6({\bf p}{\bf q})\left(\frac{p}{q}+\frac{q}{p}\right)+6\frac{({\bf p}{\bf q})^2}{pq}\Bigr]-
\frac{9m_1}{m_2F_2(0)}\bigl[pq-\frac{{\bf p}{\bf q})^2}{pq}\bigr],
\end{equation*}
Further integration over the momentum variables and spectral parameter $\xi$ can be performed 
analytically using \eqref{eq53}. But the answer for hyperfine splitting in the energy spectrum 
is more conveniently written in the integral form over $\xi$:
\begin{equation}
\label{eq60}
\Delta E^{hfs}_{vp}(2^7P_{3/2}-2^5P_{3/2})=\frac{\alpha^2(Z\alpha)^3\mu^3\mu_N}{135\pi m_1m_p}
\int_1^\infty\frac{\rho(\xi)d\xi}{(a+2)^4}\Bigl[
16+20\frac{m_1}{m_2}+a(32+40\frac{m_1}{m_2})+
\end{equation}
\begin{equation*}
15a^2\frac{m_1}{m_2}-a_\mu(4+8a+15a^2)-\frac{15m_1}{2m_2F_2(0)}(4+8a+3a^2)\Bigr],
\end{equation*}
\begin{equation}
\label{eq61}
\Delta E^{hfs}_{vp}(2^5P_{3/2}-2^3P_{3/2})=\frac{2}{3}\Delta E^{hfs}_{vp}(2^7P_{3/2}-2^5P_{3/2})=
2\Delta E^{hfs}_{vp}(2^3P_{3/2}-2^1P_{3/2}),
\end{equation}
\begin{equation}
\label{eq62}
\Delta E^{hfs}_{vp}(2^5P_{1/2}-2^3P_{1/2})=\frac{2\alpha^2(Z\alpha)^3\mu^3\mu_N}{81\pi m_1m_p}
\int_1^\infty\frac{\rho(\xi)d\xi}{(a+2)^4}\Bigl[16+32a+18a^2+
\end{equation}
\begin{equation*}
2\frac{m_1}{m_2}(4+8a+3a^2)+a_\mu(8+16a+12a^2)-\frac{3m_1}{m_2F_2(0)}(4+8a+3a^2)\Bigr].
\end{equation*}
In the second
order of perturbation theory we also have the VP contribution of order $\alpha^5$. In this
case one perturbation potential is determined by \eqref{eq36} and other perturbation is the 
vacuum polarization correction to the Coulomb potential \eqref{eq52}. For the calculation of this type
correction it is convenient to use coordinate representation in which the Coulomb Green function has
the form \cite{hameka}:
\begin{equation}
\label{eq63}
G_{2P}(\boldsymbol r,\boldsymbol r')=-\frac{\mu^2(Z\alpha)}{36x_1^2x_2^2}\biggl( \frac{3}{4\pi}
\boldsymbol n \boldsymbol n'\biggl) e^{-(x_1+x_2)/2}g(x_1,x_2),
\end{equation}
\begin{gather*}
g(x_1,x_2)=24x_<^3+36x_<^3x_>+36x_<^3x_>^2+24x_>^3+36x_<x_>^3+36x_<^2x_>^3+49x_<^3x_>^3-3x_<^4x_>^3-\\
-12e^{x_<}(2+x_<+x_<^2)x_>^3-3x_<^3x_>^4+12x_<^3x_>^3[-2C+Ei(x_<)-ln x_<-ln x_>],
\end{gather*}
where $C=0.5772...$ is the Euler constant, $x_1=Wr$, $x_2=Wr'$, $x_<=min(x_1,x_2)$, $x_>=max(x_1,x_2)$.
Then making the analytical integration over particle coordinates we obtain the following integral
representation for this correction:
\begin{equation}
\label{eq64}
\Delta E^{hfs}_{vp,sopt}=\frac{\alpha^5 Z^3\mu^3\mu_N}{144\pi m_1m_p}\int_1^\infty\frac{\rho(\xi)d\xi}{(1+b_1)^5}
[3+11b_1+4b_1^2+4(1+b_1)\ln(1+b_1)]\times
\end{equation}
\begin{equation*}
[\bar T_1+\frac{m_1}{m_2}\bar T_1-\frac{3m_1}{2m_2F_2(0)}\bar T_1-(1+a_\mu)\bar T_2],~~~b_1=\frac{2m_e}{W}.
\end{equation*}
The summary VP correction of order $\alpha^5$ from first and second order perturbation theory 
is presented in the Table~\ref{tb2} for separate energy levels.

Nucleus of lithium, beryllium and boron have sufficiently large size, so the structure
effects can be significant. For their estimation in order $\alpha^6$ we use an expansion of 
charge, magnetic dipole and electric quadrupole form factors:
\begin{equation}
\label{eq65}
F_1(k^2)\approx 1-\frac{1}{6}r_{E0}^2{\bf k}^2,~
F_2(k^2)\approx F_2(0)[1-\frac{1}{6}r_{M}^2{\bf k}^2],~
F_3(k^2)\approx 2[1-\frac{1}{6}r_{E0}^2{\bf k}^2]-2G_{E2}(0)[1-\frac{1}{6}r_{E2}^2{\bf k}^2],
\end{equation}
and take into account terms proportional to charge $r_E$, magnetic dipole $r_{M1}$ and electric quadrupole $r_{E2}$
radii. Then in momentum representation the potentials giving the splitting of P-states are the following:
\begin{equation}
\label{eq66}
\Delta V_{str}(2^7P_{3/2}-2^5P_{3/2})=\frac{Z\alpha}{45m_1m_2}
\Bigl\{-r_{E0}^2\frac{15m_1}{4m_2}[pq-\frac{({\bf p}{\bf q})^2}{pq}]
+G_{E2}(0)r_{E2}^2\frac{m_1}{2m_2}[3pq+\frac{({\bf p}{\bf q})^2}{pq}]+
\end{equation}
\begin{equation*}
F_2(0)r_{M1}^2
\bigl[2pq-6\frac{({\bf p}{\bf q})^2}{pq}+\frac{5m_1}{2m_2}\bigl(pq-\frac{({\bf p}{\bf q})^2}{pq}\bigr)-
\frac{a_\mu}{2}\bigl(pq+7\frac{({\bf p}{\bf q})^2}{pq}\bigr)\bigr]
\Bigr\},
\end{equation*}
\begin{equation}
\label{eq67}
\Delta V_{str}(2^5P_{3/2}-2^3P_{3/2})=\frac{2Z\alpha}{135m_1m_2}
\Bigl\{-r_{E0}^2\frac{15m_1}{4m_2}[pq-\frac{({\bf p}{\bf q})^2}{pq}]
-G_{E2}(0)r_{E2}^2\frac{m_1}{2m_2}[3pq+\frac{({\bf p}{\bf q})^2}{pq}]+
\end{equation}
\begin{equation*}
F_2(0)r_{M1}^2
\bigl[2pq-6\frac{({\bf p}{\bf q})^2}{pq}+\frac{5m_1}{2m_2}\bigl(pq-\frac{({\bf p}{\bf q})^2}{pq}\bigr)-
\frac{a_\mu}{2}\bigl(pq+7\frac{({\bf p}{\bf q})^2}{pq}\bigr)\bigr]
\Bigr\},
\end{equation*}
\begin{equation}
\label{eq68}
\Delta V_{str}(2^3P_{3/2}-2^1P_{3/2})=\frac{Z\alpha}{135m_1m_2}
\Bigl\{-r_{E0}^2\frac{15m_1}{4m_2}[pq-\frac{({\bf p}{\bf q})^2}{pq}]
-G_{E2}(0)r_{E2}^2\frac{m_1}{4m_2}[3pq+\frac{({\bf p}{\bf q})^2}{pq}]+
\end{equation}
\begin{equation*}
F_2(0)r_{M1}^2
\bigl[2pq-6\frac{({\bf p}{\bf q})^2}{pq}+\frac{5m_1}{2m_2}\bigl(pq-\frac{({\bf p}{\bf q})^2}{pq}\bigr)-
\frac{a_\mu}{2}\bigl(pq+7\frac{({\bf p}{\bf q})^2}{pq}\bigr)\bigr]
\Bigr\},
\end{equation*}
\begin{equation}
\label{eq69}
\Delta V_{str}(2^5P_{1/2}-2^3P_{1/2})=\frac{2Z\alpha}{27m_1m_2}
\Bigl\{-r_{E0}^2\frac{3m_1}{2m_2}[pq-\frac{({\bf p}{\bf q})^2}{pq}]+
\end{equation}
\begin{equation*}
F_2(0)r_{M1}^2
\bigl[2pq+\frac{m_1}{m_2}pq-\frac{m_1}{m_2}\frac{({\bf p}{\bf q})^2}{pq}+
a_\mu\bigl(pq+\frac{({\bf p}{\bf q})^2}{pq}\bigr)\bigr]
\Bigr\}.
\end{equation*}
The calculation of remaining momentum integrals in \eqref{eq2} gives 
$<pq>=3/8$, $<\frac{({\bf p}{\bf q})^2}{pq}>=1/8$ and shifts of the
energy levels $2^{2F+1}P_{J}$. Corresponding numerical results are presented in Table~\ref{tb2}.
To obtain them we set approximately $r_{E0}=r_{M1}$ and omit quadruple radius $r_{E2}$.

Among other important corrections of order $\alpha^6$ we can distinguish relativistic corrections
which can have large numerical values due to the factor $Z^5$. 
They can be calculated by means of the Dirac equation \cite{breit,rose}.
The expectation value of hyperfine part of the Dirac Hamiltonian can be expressed in terms of reduced
matrix elements by means of the Wigner-Eckart theorem \cite{apm2015}:
\begin{equation}
\label{eq70}
\Delta E^{hfs}_{rel}=e g_N\tilde\mu_N (-1)^{s_2+j'-F}W(j s_2 j' s_2;F 1)\Braket{s_2||\boldsymbol{s_2}||s_2}\Braket{j'||\frac{[\boldsymbol{r}\times \boldsymbol{\alpha}]}{r^3}||j},
\end{equation}
where $\tilde\mu_N$ is the nuclear magneton, $W(j s_2 j' s_2;F 1)$ are the Racah coefficients. 
The calculation of reduced matrix elements for the P-states with 
nucleus spin $3/2$ gives the following results:
\begin{equation}
\label{eq71}
E^{hfs}_{rel}(2P_{1/2})=\frac{47 Z^5\alpha^6\mu_Nm_1^2}{1296m_p}\Bigl[F(F+1)-\frac{9}{2}\Bigr],
\end{equation}
\begin{equation}
\label{eq72}
E^{hfs}_{rel}(2P_{3/2})=\frac{7 Z^5\alpha^6\mu_Nm_1^2}{6480m_p}\Bigl[F(F+1)-\frac{15}{2}\Bigr],
\end{equation}
Indeed, numerically these corrections are important to achieve high accuracy of the total result
(see Table~\ref{tb2}).

\begin{table}[htbp]
\caption{Nondiagonal matrix elements in the hyperfine structure of P-wave states of muonic lithium, 
beryllium and boron. The first, second and third lines correspond to lithium, beryllium and boron.}
\label{tb3}
\bigskip
\begin{tabular}{|c|c|c|}   \hline
Contribution to HFS       & $2^3P_{1/2,3/2}$, meV& $2^5P_{1/2,3/2}$, meV\\   \hline
Leading order $\alpha^4$        & -30.2419           &     40.2378      \\
correction   &  25.8382    &  -35.3158  \\   
   &  -116.2435    &  154.8861    \\    \hline
Leading order $\alpha^4$   &  -111.4813   &  -149.5678    \\
quadrupole correction & 347.9783   &  466.8619  \\  
    &   526.4559   &      706.3147 \\ \hline
Vacuum polarization      &  -0.0548  &   0.1223     \\
correction of order $\alpha^5$  & 0.0566  & -0.1272    \\ 
    &   -0.2842   &  0.6341     \\  \hline
Quadruple and vacuum    & -0.1901  & -0.2551    \\
polarization correction   &   0.7283  &   0.9771    \\
of order $\alpha^5$  &  1.2681   &  1.7013      \\    \hline
Relativistic correction    &   -0.0083           & 0.0111            \\
of order $\alpha^6$ &  0.0126 &  -0.0169  \\  
   &  -0.0879  & 0.1179  \\    \hline
Vacuum polarization      & -0.0008 &  0.0018  \\
correction of order $\alpha^6$  & 0.0008   &  -0.0018   \\ 
  &  -0.0037   &   0.0082   \\   \hline
Summary contribution  &  -141.9772 &  -109.4499  \\  
   &  374.6148   &   432.3773   \\     
   &      411.1047    &     863.6623   \\      \hline
\end{tabular}
\end{table}

\section{Nondiagonal matrix elements}

Up to this point we have considered diagonal matrix elements between different states
$2^{(2F+1)}P_j$. But the one-photon interaction Hamiltonian leads to the mixing of states 
$2^3P_{1/2}$, $2^3P_{3/2}$ and $2^5P_{1/2}$, $2^5P_{3/2}$. To calculate transitions
between these states we use developed in previous sections formalism of projection operators.
Then the general structure of transition amplitudes between states $2^3P_{1/2}$, $2^3P_{3/2}$ and
$2^5P_{1/2}$, $2^5P_{3/2}$ is the following:
\begin{equation}
\label{eq73}
<2^3P_{1/2}|T_{1\gamma}|2^3P_{3/2}>=Z\alpha \sqrt{3}n_q^\delta n_p^\omega
Tr\Bigl\{[v_{\beta_1}(0)\bar\psi(0)]_{j=1/2}^{F=1}
(\gamma_\delta-v_\delta)\gamma_5\frac{(\hat q_1+m_1)}{2m_1}
\Gamma_\mu\frac{(\hat p_1+m_1)}{2m_1}\times
\end{equation}
\begin{equation*}
[\psi_\omega(0)\bar v_{\alpha_1}(0)]^{F=1}_{j=3/2}
\frac{(\hat p_2-m_2)}{2m_2}\Gamma_{\alpha\beta}^\nu\frac{(\hat q_2-m_2)}{2m_2}
\Bigr\}D_{\mu\nu}(k)L_{\alpha\alpha_1}L_{\beta\beta_1},
\end{equation*}
\begin{equation}
\label{eq74}
<2^5P_{1/2}|T_{1\gamma}|2^5P_{3/2}>=Z\alpha \sqrt{3} n_q^\delta n_p^\omega
Tr\Bigl\{[v_{\beta_1}(0)\bar\psi(0)]_{j=1/2}^{F=2}
(\gamma_\delta-v_\delta)\gamma_5\frac{(\hat q_1+m_1)}{2m_1}
\Gamma_\mu\frac{(\hat p_1+m_1)}{2m_1}\times
\end{equation}
\begin{equation*}
[\psi_\omega(0)\bar v_{\alpha_1}(0)]^{F=2}_{j=3/2}
\frac{(\hat p_2-m_2)}{2m_2}\Gamma_{\alpha\beta}^\nu\frac{(\hat q_2-m_2)}{2m_2}
\Bigr\}D_{\mu\nu}(k)L_{\alpha\alpha_1}L_{\beta\beta_1},
\end{equation*}
Using projection operators \eqref{eq11}, \eqref{eq28}, \eqref{eq30}, \eqref{eq32} we obtain
contributions to the energy spectrum of order $\alpha^4$:
\begin{equation}
\label{eq75}
E^{F=1}(j=1/2;j=3/2)=-\frac{\sqrt{5}\alpha^4 Z^3\mu^3\mu_N}{216m_1m_p}\left[1+\frac{2m_1}{m_2}-
a_\mu+\frac{m_1}{m_2F_2(0)}\right],
\end{equation}
\begin{equation}
\label{eq76}
E^{F=2}(j=1/2;j=3/2)=\frac{\alpha^4 Z^3\mu^3\mu_N}{72m_1m_p}\left[1+\frac{2m_1}{m_2}-
a_\mu-\frac{3m_1}{m_2F_2(0)}\right].
\end{equation}
In one-photon approximation there exists also another correction of order $\alpha^4$  
connected with the quadrupole electric form factor for the nucleus with spin 3/2. We present 
the correction to the quadrupole interaction in separate equation as \eqref{eq51}:
\begin{equation}
\label{eq77}
\Delta E_Q^{hfs}(j=1/2;j=3/2)=\frac{\alpha Q(\mu Z\alpha)^3}{60}\left[\delta_{F2}+\frac{\sqrt{5}}{3}\delta_{F1}\right].
\end{equation}
The contributions of leading order \eqref{eq75}, \eqref{eq76}, \eqref{eq77} must be supplemented
by the same vacuum polarization corrections which are calculated above for diagonal matrix elements.
Let us write them in integral form over spectral parameter:
\begin{equation}
\label{eq78}
E^{F=1}(j=1/2;j=3/2)=-\frac{\sqrt{5}\alpha^5 Z^3\mu^3 \mu_N}{162\pi m_1m_p}\int_1^\infty
\frac{\rho(\xi)d\xi}{(a+2)^4}\Bigl[4+8a+9a^2+
\end{equation}
\begin{equation*}
2\frac{m_1}{m_2}(4+8a+3a^2)-
a_\mu(4+8a-3a^2)+\frac{m_1}{m_2F_2(0)}(4+8a+3a^2)\Bigr],
\end{equation*}
\begin{equation}
\label{eq79}
E^{F=2}(j=1/2;j=3/2)=\frac{5\alpha^5 Z^3\mu^3 \mu_N}{162\pi m_1m_p}\int_1^\infty
\frac{\rho(\xi)d\xi}{(a+2)^4}
\Bigl[4+8a+9a^2+
\end{equation}
\begin{equation*}
2\frac{m_1}{m_2}(4+8a+3a^2)-
a_\mu(4+8a-3a^2)-\frac{3m_1}{m_2F_2(0)}(4+8a+3a^2)\Bigr].
\end{equation*}
The contribution of vacuum polarization to quadrupole interaction also can be obtained with
the use of Eqs.\eqref{eq73}-\eqref{eq74} in the form:
\begin{equation}
\label{eq80}
E_{Q,vp}(j=1/2;j=3/2)=\frac{\alpha^5Z^3\mu^3Q}{45\pi}
\int_1^\infty\frac{\rho(\xi)d\xi}{(a+2)^4}[4+8a+5a^2]\left[\delta_{F2}+\frac{\sqrt{5}}{3}\delta_{F1}\right].
\end{equation}
We can not neglect by relativistic corrections to nondiagonal matrix elements. General expression \eqref{eq70}
for its calculation contains the following radial integral with the Dirac wave functions \cite{wgr}:
\begin{equation}
\label{eq81}
R_{\frac{1}{2}\frac{3}{2}}=\int_0^\infty[g_{1/2}(r)f_{3/2}(r)+g_{3/2}(r)f_{1/2}(r)]dr=
\frac{(Z\alpha)^3}{48}\left[1+\frac{9(Z\alpha)^2}{16}\right]+O((Z\alpha)^7),
\end{equation}
where the indexes $1/2$ and $3/2$ designate the total muon momentum.
In final form relativistic corrections to nondiagonal matrix elements are determined by
following expressions:
\begin{equation}
\label{eq82}
E_{rel}^{F=1}(j=1/2;j=3/2)=-\frac{\sqrt{5}\alpha^6Z^5m_1^2\mu_N}{384m_p},
\end{equation}
\begin{equation}
\label{eq83}
E_{rel}^{F=2}(j=1/2;j=3/2)=\frac{\alpha^6Z^5m_1^2\mu_N}{128m_p}.
\end{equation}
All equations of this section \eqref{eq75}-\eqref{eq83} are used for obtaining numerical
results which are presented in Table~\ref{tb3}.

\section{Conclusion}

Hyperfine structure of the P-wave energy spectrum plays an important role for precise
calculation of transition frequencies between energy levels. We study  
the hyperfine structure of energy levels in muonic ions of lithium, beryllium and boron 
on the basis quasipotential method in quantum electrodynamics. Our calculation contains
the leading order $\alpha^4$ contribution and corrections of orders $\alpha^5$ and $\alpha^6$
to the vacuum polarization, nucleus structure, quadrupole interaction and relativism.
These corrections have significant numerical value because of the factor proportional to
the nucleus charge. The used formalism allows to calculate corrections in analytical form
what is demonstrated in different places of the work. In a number of cases we preserve 
integral representation over some spectral parameter for corrections to the vacuum polarization 
because it is more compact.
Numerical results are written in Tables~\ref{tb2},\ref{tb3},\ref{tb4}.

This work continues our investigation of P-wave part of the energy spectrum in light muonic 
atoms which was begun in \cite{apm2015,apm2010,apm2008,apmepj}. Taking nucleus of spin 3/2 we obtain new results
in the study of hyperfine structure which are as follows:

1. We are developing the method of projection operators for spin nuclei 3/2 in the momentum 
representation, which allows us to construct the interaction operator of the muon and nucleus 
for various states. The results of calculating the contributions to the energy spectrum 
by this method are consistent with the calculation performed in the framework of the coordinate 
representation.

2. The corrections of the fifth-order in $\alpha$ on vacuum polarization are calculated in the 
hyperfine structure of the P-states, including the quadrupole interaction.

3. An estimate of the sixth-order $\alpha$ contributions to the structure of the nucleus 
and the vacuum polarization is obtained. Relativistic corrections are also taken into account 
in the framework of the relativistic Dirac equation.

In last years the calculation of different corrections to fine and hyperfine structure 
of muonic atoms was done by several groups. Because of large number of the works we give 
only references on review articles \cite{crema1,pohl2013,bacca,egs,borie,sgk} 
which contain many other references regarding 
to the problem. The hyperfine structure of P-states in muonic lithium, beryllium 
and boron was not calculated directly in these papers. There is the only paper 
\cite{drake2} in which  the basic analytical formula for the contribution of order $O(\alpha^4)$ 
without recoil was presented.
We have improved previous estimates of P-energy levels obtained by authors \cite{drake2}  
taking into account new corrections.
For definiteness, we present numerical results with an accuracy of 4 digits after the decimal point. 
Errors in the determination of fundamental physical constants are negligible when obtaining 
the final results. The accuracy of the theoretical calculation is determined by the omitted 
corrections of a higher order in $\alpha$.
On the whole, it can be said that we obtained the values of the hyperfine splittings of the 
P-levels with an accuracy of 0.001 meV, which can be used as a guide for comparison 
with future experimental data.

Summing  all diagonal and off-diagonal matrix elements in the case of muonic lithium 
we obtain the following energy matrix
\begin{widetext}
\begin{equation}
\label{eq:2P_matrix}
\small{
\begin{aligned}
M ~
  &= & & \bordermatrix{
              &2^3P_{1/2}&2^5P_{1/2}&2^1P_{3/2}&2^3P_{3/2}&2^5P_{3/2}& 2^7P_{3/2}  \\[1ex]
2^3P_{1/2} & -132.5885 &  0          & 0   &  -141.9772         &  0  &   0     \\[1ex]
2^5P_{1/2} &  0          &  79.5517&  0           & 0 &  -109.4499  &  0      \\[1ex]
2^1P_{3/2} &  0  &  0          &  482.5174&  0          &  0  &   0     \\[1ex]
2^3P_{3/2} &  -141.9772          & 0 &  0           &  653.9634 0   &     0  \\[1ex]
2^5P_{3/2} &  0          &  -109.4499          &  0           &  0          &  845.9732&  0   \\[1ex]
2^7P_{3/2} &  0          &  0          &  0           &  0          &  0 &  760.7254  \\[1ex]
} ~ \mathrm{meV.}
\end{aligned}
}
\end{equation}
\end{widetext}

\begin{table}[htbp]
\caption{Hyperfine structure of P-states in muonic ions of lithium, beryllium and boron.}
\label{tb4}
\bigskip
\begin{tabular}{|c|c|c|c|}   \hline
State       & $(\mu Li)^{2+}$ energy, meV & $(\mu Be)^{3+}$ energy, meV & $(\mu B)^{4+}$ energy, meV  \\ \hline
$2^3P_{1/2}$ & -157.432       &   57.912   & -543.912 \\ \hline
$2^5P_{1/2}$  &64.228       &  -154.353   & 163.415  \\ \hline
$2^1P_{3/2}$  & 482.517      &   3038.062   &   6413.094 \\ \hline
$2^3P_{3/2}$  & 678.806         &  2607.268  & 5812.673   \\ \hline
$2^5P_{3/2}$   & 861.297        &  2129.583   &  5387.603  \\ \hline
$2^7P_{3/2}$   &  760.725    &    2457.571 &  6199.288 \\ \hline
\end{tabular}
\end{table} 

Its diagonalization leads directly to the position of the energy levels $2P$ (see Table~\ref{tb4}) 
and hyperfine splitting intervals for muonic lithium which can be measured in the experiment. 
In the same way we obtained hyperfine structure of two other ions of muonic beryllium and boron
which are written also in Table~\ref{tb4}.

\begin{acknowledgments}
The work is supported by Russian Science Foundation (grant No. RSF 18-12-00128) and
Russian Foundation for Basic Research (grant No. 18-32-00023) (F.A.M.).
\end{acknowledgments}

\end{document}